\documentclass[twocolumn,prd,preprintnumbers,tightenlines,footinbib,superscriptaddress]{revtex4}

\pdfoutput=1
\usepackage[english]{babel}
\usepackage{amsmath,amssymb,amsfonts}
\usepackage{graphicx}
\usepackage[normalem]{ulem}
\usepackage{slashed}
\usepackage{booktabs}
\usepackage{hyperref}
\usepackage{color}
\usepackage[sort&compress]{natbib}
\usepackage{tikz}
\usepackage{comment}

\usepackage{cleveref}
\newcounter{qnumber}

\newcommand{\N}{N_c}
\newcommand{\Ntil}{\widetilde{N}_c}
\newcommand{\F}{N_f}
\newcommand{\Mtil}{\widetilde{M}}
\newcommand{\Bbar}{\bar{B}}
\newcommand{\bbar}{\bar{b}}
\newcommand{\id}{\textbf{1}}
\newcommand{\Kah}{K\"ahler }
\newcommand{\Lambdatil}{\widetilde{\Lambda}}
\DeclareMathOperator{\Tr}{Tr}

\begin{document}

\title{A Guide to AMSB QCD}

\author{Csaba Cs\'aki}
\email[]{ccsaki@gmail.com}
\affiliation{Department of Physics, LEPP, Cornell University, Ithaca, NY 14853, USA}

\author{Andrew Gomes}
\email[]{awg76@cornell.edu}
\affiliation{Department of Physics, LEPP, Cornell University, Ithaca, NY 14853, USA}

\author{Hitoshi Murayama}
\email[]{hitoshi@berkeley.edu}
\affiliation{Department of Physics, University of California, Berkeley, CA 94720, USA}
\affiliation{Kavli Institute for the Physics and Mathematics of the
  Universe (WPI), University of Tokyo,
  Kashiwa 277-8583, Japan}
\affiliation{Ernest Orlando Lawrence Berkeley National Laboratory, Berkeley, CA 94720, USA}

\author{Bea Noether}
\email[]{bea\_noether@berkeley.edu}
\affiliation{Department of Physics, University of California, Berkeley, CA 94720, USA}

\author{Digvijay Roy Varier}
\email[]{digvijayroyvarier@berkeley.edu}
\affiliation{Department of Physics, University of California, Berkeley, CA 94720, USA}

\author{Ofri Telem}
\email[]{t10ofrit@gmail.com}
\affiliation{Racah Institute of Physics, Hebrew University of Jerusalem, Jerusalem 91904, Israel}

\begin{abstract} 
We present a careful study of the chiral symmetry breaking minima and the baryonic directions in supersymmetric QCD ($SU(\N)$ with $\F$ flavors) perturbed by Anomaly Mediated Supersymmetry Breaking (AMSB). For the s-confining case of $\F = \N + 1$ and most of the free-magnetic phase ($\F \leq 1.43 \N$) we find that naive tree level baryonic runaways are stabilized by loop effects. Runaways are present, however, for the upper end of the free magnetic phase ($\F \gtrsim 1.43 \N$) and into conformal window, signaling the existence of incalculable minima at large field values of  ${\cal O} (\Lambda)$. Nevertheless, the chiral symmetry breaking points are locally stable, and are expected to continuously connect to the vacua of QCD for large SUSY breaking. The case of $\F = \N$ requires particular care due to the inherently strongly coupled nature of the quantum modified moduli space. Due to the incalculability of critical \Kah potential terms, the stability of the chiral symmetry breaking point along baryonic directions cannot be determined for $\F=\N$. With the exception of this case, all theories to which AMSB can be applied ($\F < 3 N_c$) possess stable chiral symmetry breaking minima, and all theories with $\F \lesssim 1.43 \N$ (aside from $\F = \N$) are protected from runaways to incalculable minima.
\end{abstract}

\maketitle

\section{Introduction}

One of the greatest challenges facing particle physics and quantum field theory (QFT) is to establish the phase structure of strongly coupled gauge theories. In particular, that of ordinary Quantum Chromodynamics (QCD), corresponding to the observed color confinement with chiral symmetry breaking. While eventually we expect lattice simulations to settle this issue, at least for non-chiral theories, progress has been quite slow and there are very few analytic tools at our disposal. One possible approach is to use the exact results and phase structures of the supersymmetric (SUSY) versions of these theories (SQCD) worked out by Seiberg and others in the 1990's~\cite{Affleck:1983mk,Seiberg:1994bz,Seiberg:1994pq}, and to add small SUSY breaking perturbations~\cite{Evans:1995ia,Aharony:1995zh,Evans:1995rv,DHoker:1996xdz,Alvarez-Gaume:1996vlf,Alvarez-Gaume:1996qoj,Evans:1996hi,Konishi:1996iz,Alvarez-Gaume:1997bzm,Evans:1997dz,Alvarez-Gaume:1997wnu,Cheng:1998xg,Martin:1998yr,Arkani-Hamed:1998dti,Luty:1999qc,Abel:2011wv,Cordova:2018acb}. The exact mapping of SUSY breaking perturbations from the UV theory to its IR manifestation has been done by linking the SUSY breaking either to holomorphic quantities \cite{Cheng:1998xg}, or to conserved and anomalous currents \cite{Arkani-Hamed:1998dti,Luty:1999qc,Abel:2011wv,Cordova:2018acb}. While being successful in mapping UV supersymmetry breaking to the IR, in many previous attempts at studying the vacuum structure of softly broken SQCD the eventual IR phase was incalculable due to runaways and/or dependence on unknown \Kah terms. For this reason, they were of limited predictivity.

A systematic study of the phases of SUSY $SU(N_c)$ gauge theories perturbed via anomaly mediated supersymmetry breaking (AMSB) was initiated in~\cite{Murayama:2021xfj}, and many new results using this method have already been obtained. These include novel symmetry breaking patterns for chiral gauge theories \cite{Csaki:2021xhi,Csaki:2021aqv,Kondo:2022lvu}, the description of confinement in $SO(N_c)$ theories via monopole condensation \cite{Csaki:2021jax}, and the phase structure of the $SO(N_c)$ theories while varying the number $\F$ of matter fields in the vector representation \cite{Csaki:2021xuc}. The result of the $SO(N_c)$ analysis was that the various exotic SUSY phases collapse as a result of SUSY breaking, and one is left only with the expected confining and chiral symmetry breaking phase. Interestingly, the analysis of the basic $SU(N_c)$ theories with $\F$ flavors of quarks turns out to be the most subtle one. A QCD-like vacuum with a chiral symmetry breaking pattern of the form $SU(\F)_L \times SU(\F)_R \to SU(\F)_D$ has been identfied in~\cite{Murayama:2021xfj,Murayama:2021bndrv} which appears to be the global minimum for at least $\F < N_c$. However, the $\F \geq N_c$ cases are complicated by the appearance of baryonic directions, which in many cases appear to cause a runaway behavior. 

The aim of this paper is to carefully examine the $SU(N_c)$ theories for $\F\geq N_c$ and, in particular, the fate of the baryonic directions \footnote{We thank Nathaniel Craig for emphasizing to us the importance of properly analyzing the baryonic directions and their potential runaways for establishing the correct vacuum of the theory.}, thus establishing the phase structure of the $SU(N_c)$ theories, when it is calculable. We will show that for the special case of $\F=N_c+1$ the potential baryonic runaway is stabilized by a 2-loop AMSB effect, while for $\F=N_c$ the theory is incalculable along these directions, and one can not conclusively decide if the baryonic runaways are lifted or not. The lower end $N_c+1 < \F \lesssim 1.43 N_c$ of the free magnetic phase will again have the baryonic runaways lifted via 2-loop AMSB, and one ends up with stable, calculable vacuum with chiral symmetry breaking. Such a ``QCD-like" vacuum with chiral symmetry breaking exists for any number of flavors as long as $\F < 3 N_c$ (with the possible exception of $\F=N_c$ where its stability cannot be determined).

In contrast, for $\F \gtrsim 1.43 N_c$ the baryonic directions will indeed contain runaways. We stress that these runaways do not invalidate the theory since they are cured once the field vacuum expectation values (VEVs) are of $\mathcal{O}(\Lambda_\text{QCD})$. Here the IR description breaks down and one must return to the UV description, where the theory is stabilized by AMSB. Instead, they merely signal that the global minimum lies in the incalculable region where field VEVs are of $\mathcal{O}(\Lambda_\text{QCD})$. In addition, the QCD-like minimum will persist as a local minimum, and one expects that as the magnitude of SUSY breaking is increased it will indeed take over as the true vacuum. Note also that baryonic runaway does not occur in $Sp$ or $SO$ gauge theories. We will discuss these cases elsewhere.

The paper is organized as follows. We first review the AMSB mechanism and then its application to the case $\F < N_c$, where chiral symmetry breaking is observed. We then successively increase the number of flavors, exhibiting chiral symmetry breaking behavior and discussing the baryonic directions, before concluding.

\section{Anomaly Mediation}

Anomaly mediation of supersymmetry breaking (AMSB)~\cite{Randall:1998uk,Giudice:1998xp} (see also \cite{ArkaniHamed:1998kj,Arkani-Hamed:1998dti} for earlier work containing some important aspects of AMSB) is parameterized by a single spurion $m$ that explicitly breaks supersymmetry in two different ways. One is the tree-level contribution based on the \Kah potential and superpotential - which is easily derived using the conformal compensator formalism \cite{Pomarol:1999ie}. It is given by
\begin{align}\label{eq:extendedAMSB}
	V_{\rm tree} = & \partial_i W g^{i j^*} \partial_j^* W^*  
	+ m^* m \left( \partial_i K g^{i j^*} \partial_j^* K - K \right) \nonumber \\
	&\quad + m \left(\partial_i W g^{i j^*} \partial_j^* K - 3 W \right) + c.c.
\end{align}
where $g^{ij}$ is the inverse of the K\"{a}hler metric $g_{ij} = \partial_i \partial_j^* K$. For simplicity we will always take $m$ to be real. Note that (\ref{eq:extendedAMSB}) breaks the $U(1)_R$ symmetry explicitly. When the \Kah potential is canonical, this reduces to the more familiar
\begin{align}
	{V}_{\rm tree} &= m \left( \varphi_{i} \frac{\partial W}{\partial \varphi_{i}} - 3 W \right)
	+ c.c. \label{eq:AMSBW}
\end{align}
When the superpotential does not include dimensionful parameters, this expression identically vanishes. 

In this case, there are the loop-level supersymmetry breaking effects from the superconformal anomaly \cite{Pomarol:1999ie}. They lead to tri-linear couplings, scalar masses, and gaugino masses,
\begin{align}
	A_{ijk} (\mu) &= - \frac{1}{2} (\gamma_{i} + \gamma_{j} + \gamma_{k})(\mu)\, m \label{eq:AMSBloop_trilin}\\
	m_{i}^{2}(\mu) &= - \frac{1}{4} \dot{\gamma}_{i}(\mu)\, m^{2} \label{eq:AMSBloop}\\
	m_{\lambda}(\mu) &= - \frac{\beta(g^{2})}{2g^{2}}(\mu)\, m.\label{eq:AMSBloop_glu}
\end{align}
Here, $\gamma_{i} = \mu\frac{d}{d\mu} \ln Z_{i}(\mu)$, $\dot{\gamma} = \mu \frac{d}{d\mu} \gamma_{i}$, and $\beta(g^{2}) = \mu \frac{d}{d\mu} g^{2}$. When the gauge theory is asymptotically free, $m_i^2>0$, stabilizing the theory against run-away behavior.

Therefore, in a theory described in the UV description by an $SU(\N)$ gauge group and $\F$ flavors such that $\F < 3 \N$, AMSB prepares exactly the state we are looking for: the squarks and gauginos become massive while the massless degrees of freedom are those of non-SUSY QCD. By the UV insensitive nature of AMSB, the expressions above can be reliably used in the dual (IR) description of the theory to determine the low-energy phase.


Here we present some expressions that will be useful later on. Suppose we have a $SU(\Ntil)$ gauge theory with gauge coupling $g$ and a superpotential $W = \lambda \Tr q_i M_{i j} \bar{q}_j$, where the $q_i$ ($\bar{q}_j$) are $\F$ flavors of (anti-)quarks and $M_{ij}$ is a gauge-singlet flavor-bifundamental meson. The anomalous dimensions are
\begin{align}
    \gamma_q &= \frac{C_F g^2}{4\pi^2} - \frac{\F \lambda^2}{8\pi^2} \\
    \gamma_M &= -\frac{\Ntil \lambda^2}{8\pi^2}
\end{align}
where $C_F = (\Ntil^2 - 1)/(2\Ntil)$ is the quadratic Casimir of the dual gauge group. For the 1-loop beta functions one has
\begin{align}
    \beta(g^2) &= -\frac{\widetilde{b} g^4}{8\pi^2} \\
    \beta(\lambda^2) &= -(\gamma_M + 2\gamma_q) \lambda^2
\end{align}
where $\widetilde{b} = 3\Ntil - \F$.

\section{$\F < \N$: ADS superpotential}

For completeness we quickly review here the results of \cite{Murayama:2021xfj} for $\F < \N$. The dynamics is described in terms of the meson fields $M_{ij}$ with the Affleck--Dine--Seiberg (ADS) superpotential
\begin{align}
	W = (N_{c} - \F) \left( \frac{\Lambda^{3N_{c}-\F}}{{\rm det} M} \right)^{1/(N_{c}-\F)}.
\end{align}

In the SUSY limit, this produces a runaway potential and hence has no ground states. When $M \gg \Lambda^{2}$, $M_{ij} = M \delta_{ij}$ describes the $D$-flat direction
\begin{align}
	Q = \bar{Q} = \left(\begin{array}{ccc} 
		1 & \cdots & 0 \\ 
		\vdots &\ddots & \vdots \\
		0 & \cdots & 1 \\ \hline
		0 & \cdots & 0 \\
		\vdots & \vdots & \vdots \\
		0 & \cdots & 0
		\end{array} \right) \phi, \qquad M = \phi^{2}		.
\end{align}
Here, $Q$ and $\bar{Q}$ are the quark/anti-quark superfields. The upper part is an $\F \times \F$ block, while the lower part is $(N_{c}-\F) \times \F$. Since the \Kah potential is canonical in the variable $\phi$, one can use (\ref{eq:AMSBW}) to obtain
\begin{align}
	V &=\left|2\F \frac{1}{\phi} 
	\left(\frac{\Lambda^{3N_{c}-\F}}{\phi^{2\F}} \right)^{1/(N_{c}-\F)} \right|^{2}
	\nonumber \\
	& - (3N_{c} - \F )
	m \left( \frac{\Lambda^{3N_{c}-\F}}{\phi^{2\F}} \right)^{1/(N_{c}-\F)}
	+ c.c.
\end{align}

Note that there is now a minimum at
\begin{align}
	M_{ij} &= \Lambda^{2} 
	\left( \frac{4\F(N_{c}+\F)}{3N_{c}-\F} \frac{\Lambda}{m} \right)^{(N_{c}-\F)/N_{c}}
	\delta_{ij} .
\end{align}
The minimum is indeed at $M \gg \Lambda^{2}$ which justifies the weakly-coupled analysis. The $SU(\F)_L \times SU(\F)_R$ flavor symmetry is dynamically broken to $SU(\F)_V$. The case of non-homogeneous values for the diagonal entries of $M$ was considered in \cite{Luzio:2022ccn}. There it was shown that the minimum is indeed found at $M_{ij} \propto \delta_{ij}$.

The massless particle spectrum consists of the Nambu--Goldstone bosons (pions) \footnote{ The case $N_{f}=1$ is special as there is no non-anomalous flavor symmetry and hence the spectrum is gapped.}. The scalar and fermion partners of the Nambu--Goldstone bosons (NGBs) have masses that grow with $m$. Naively increasing $m$ beyond $\Lambda$, the only remaining degrees of freedom are massless NGBs. This matchs the expectations of QCD with a small number of flavors. There is no sign of a phase transition and the two limits are likely continuously connected.

\section{$\F = \N$: Quantum modified constraint}

In this section we give a complete analysis of the case of the quantum modified constraint, finding that previous discussion requires modification.

The low-energy degrees are meson fields $M_{ij}$ and singlet baryon/anti-baryon fields $B$ ($\Bbar$), whose moduli space is subject to the quantum modified constraint
\begin{equation}\label{constraint}
    \det{M} - B \Bbar = \Lambda^{2\N} .
\end{equation}
We first treat the general case $\N > 2$, and treat the case $\N=2$ separately at the end.

There are two ways to frame the theory before the addition of AMSB. The first is to implement the constraint in the superpotential via a Lagrange multiplier field $X$. However due to the constraint (\ref{constraint}), the fields have VEVs of $\mathcal{O}(\Lambda)$. Therefore, higher order terms in the \Kah potential are not suppressed relative to the canonical term and the formula (\ref{eq:AMSBW}) cannot be trusted.

Instead we should perform a non-linear analysis using the constraint (\ref{constraint}). For simplicity, we will use units where $\Lambda = 1$. The moduli space contains two special points of enhanced symmetry: the meson point $M = \id, B = \Bbar = 0$ with unbroken baryon number, and the baryon point $M = 0, B = -\Bbar = 1$ with unbroken flavor symmetry. We perform AMSB around each of these points.

\subsection{Meson point}

To satisfy the constraint at the meson point we make the change of variables
\begin{equation}\label{meson_def}
    M = (1 + B \Bbar)^{1/\N} e^\Pi = \id + \frac{1}{\N} B \Bbar + \Pi + \frac{1}{2} \Pi^2 + \cdots\ 
\end{equation}
where $\Pi$ is a traceless complex matrix. In what follows we will work to quadratic order. The \Kah potential is built out of flavor invariants, e.g. $\Tr M^\dagger M$, $(\Tr M^\dagger M)^2$, $\Tr M^\dagger M M^\dagger M$, etc. Notice that they will all contribute at quadratic order in the hadron superfields. Let's examine the $\Pi$ contribution of the first term:
\begin{equation}\label{pi_gold_k}
    \Tr M^\dagger M \supset \Tr \Pi^\dagger \Pi + \frac{1}{2} \Tr \Pi^2 + \frac{1}{2} \Tr {\Pi^\dagger}^2 .
\end{equation}

A useful formula going forward will be the tree level AMSB potential corresponding to $K = \varphi^\dagger \varphi + \alpha/2 \,(\varphi^2 +  {\varphi^\dagger}^2)$. Using the general formula (\ref{eq:extendedAMSB}), we get
\begin{align}\label{realimage}
    V_{\rm AMSB} &= \alpha^2 m^2 \varphi^\dagger \varphi + \frac{\alpha}{2} m^2 (\varphi^2 + {\varphi^\dagger}^2)\\
    &= (\alpha^2 + \alpha)m^2 (\text{Re}\, \varphi)^2 + (\alpha^2 - \alpha)m^2 (\text{Im}\, \varphi)^2 . \nonumber
\end{align}

Setting $\alpha=1$ corresponds to the \Kah potential for each component of $\Pi$ in \eqref{pi_gold_k}, so that the $\text{Im}\, \Pi$ are the massless pions, the Goldstone bosons of broken chiral flavor symmetry. Goldstone's theorem ensures that all meson flavor invariants of the \Kah potential will give contributions proportional to the right-hand-side of (\ref{pi_gold_k}). Moreover, they will (in aggregate) come with a positive sign in order for the $\Pi$ to have a physical kinetic term. Thus, the $\text{Re}\, \Pi$ will have a positive mass, stabilizing this direction.

Turning to the baryons, things are not as clear. The most general form of the \Kah potential at quadratic order is
\begin{equation}
    K \supset \alpha (B^\dagger B + \Bbar^\dagger \Bbar) + \frac{\beta}{2} (B \Bbar + c.c.)
\end{equation}
where this includes contributions (\ref{meson_def}) from meson field traces. We cannot know the ratio $\beta/\alpha$ and thus are unable to determine whether the meson point is stable with respect to baryonic runaway to an incalculable minimum.

\subsection{Baryon point}

Here we parameterize the baryon and anti-baryon with a single complex field $b$:
\begin{align}
    B &= (1 - \det M)^{1/2} e^b \\
    \Bbar &= -(1 - \det M)^{1/2} e^{-b} .
\end{align}

Like at the meson point, we expect to find a Goldstone boson, now from spontaneously broken baryon number. Consider for example the \Kah potential terms
\begin{equation}
    B^\dagger B + \Bbar^\dagger \Bbar = 2 + (b + b^\dagger)^2 + \cdots .
\end{equation}
Again using (\ref{realimage}), we identify $\text{Im}\, b$ as the Goldstone boson, while $\text{Re}\, b$ has positive mass. Regarding the mesons however, only the quadratic term must come with a positive sign (to give positive kinetic term). The coefficients of all higher order flavor invariants in the \Kah potential are unknown. With the application of (\ref{eq:extendedAMSB}), these will ultimately determine if the baryon point is stable once AMSB is turned on.

In summary, we can say very little about the behavior of AMSB-deformed QCD in the singular case when $\F = \N$. Neither global nor local minima can be identified, though based on the behavior of theories with more or fewer flavors we can conjecture a chiral symmetry breaking minimum at the meson point. This ambiguity can be traced to the quantum modified constraint, making the theory inherently strongly coupled.

\subsection{$\N = 2$}

When $\N=2$, the quarks and anti-quarks belong to the same representation of the gauge group. Thus, the flavor symmetry is enhanced to $SU(4)$, with the meson $M$ transforming in the anti-symmetric representation. This meson can be decomposed into the meson, baryon, and anti-baryon of the unenhanced flavor symmetry. The quantum modified constraint becomes $M^a M^a = 1$, with $a = 1, \dots , 6$, meaning the moduli space has 5 complex dimensions. The constraint breaks the flavor symmetry to $Sp(4)$, leading to 5 Goldstone modes. Due to the kinetic term positivity arguments made above, their scalar partners have positive mass.

Thus, the enhanced symmetry causes the chiral symmetry breaking minimum to be stable in the case of $\N=2$. Similar results were found in \cite{Luty:1999qc}. Note that $\N=2$ is a special case of the $Sp$ gauge theories that will be discussed elsewhere.

\section{$\F = \N + 1$: S-Confinement}

For this case we find a stable chiral symmetry breaking minimum, and demonstrate that there are no runaway directions. At the leading order we take a canonical \Kah potential for low energy fields $B$, $\Bbar$ and $M$, which is justified when $B,\Bbar,M \ll \Lambda$ where the theory is weakly coupled. The superpotential is
\begin{equation}\label{eq:sconf}
    W = \alpha B M \Bbar - \beta \det M\ 
\end{equation}
where we are again working in $\Lambda = 1$ units and $\alpha$ and $\beta$ are unknown order one numbers used to make the \Kah canonical. The potential obtained is
\begin{align}
	V_{\rm SUSY} &= \alpha^2 (|(M\Bbar)_a|^2 + |(BM)_a|^2) \nonumber\\
	&\quad  + |\alpha \Bbar_a B_b - \beta \det{M} (M^{-1})_{a b} |^2 \\
	V_{\rm AMSB} &= - (\N-2) \beta m \det{M} + c.c.
\end{align}

Seeking the minimum of this potential, we look along the direction
\begin{equation}
    B = \left(\begin{array}{c} 
		b \\
		0 \\
		\vdots \\ 0
		\end{array} \right)
		,
    \Bbar = \left(\begin{array}{c} 
		\bbar \\
		0 \\
		\vdots \\ 0
		\end{array} \right)
		,
	M = \left(\begin{array}{cccc} 
		x & & &\\
		& v & &\\
		& & \ddots &\\
		& & & v
		\end{array} \right) .
\end{equation}
Using flavor rotations the baryon and anti-baryon take this form without loss of generality. They break the flavor symmetry to $SU(\N)_L \times SU(\N)_R$, justifying the inhomogeneous diagonal VEVs of $M$. For fixed $\det M$, any off-diagonal terms would simply increase $V_\text{SUSY}$, justifying their omission. Finally, given that we are taking $m$ real, it is enough to look for minima with all fields real.

Using the fact that for fixed $b \bbar$, the quantity $b^2 + \bbar^2$ is minimized when $b = \bbar$, the potential is
\begin{align}\label{eq:sconf_pot}
    V &= 2\alpha^2 x^2 b^2 + (\alpha b^2 - \beta v^{\N})^2 + \N \beta^2 x^2 v^{2(\N-1)} \nonumber \\
    &\quad - 2 (\N-2)\beta m x v^{\N} .
\end{align}
Again we treat the general case $\N > 2$ first, and the case $\N=2$ separately afterwards.

The crucial observation implicit above is that the baryon fields do not acquire tree-level SUSY breaking whose mass originates from AMSB and they do not induce threshold corrections when they are integrated out, called ``non-decoupling effects'' in \cite{Pomarol:1999ie}.

\subsection{Baryon number conserving direction, $b=0$}

For the baryon number conserving direction $b = 0$, one finds a minimum
\begin{equation}\label{eq:sconf_chisb}
    v = x = \left( \frac{(\N-2)m}{\N \beta} \right)^{\frac{1}{\N-1}} , V_{\rm min} = -\mathcal{O}(m^{2\N/(\N-1)}) .
\end{equation}

This is the chiral symmetry breaking minimum that we hope to be continuously connected to that of non-SUSY QCD. First we must check that it is not disturbed by loop effects coming from the marginal Yukawa term in (\ref{eq:sconf}). The baryons acquire a mass $\alpha v$, and integrating them out and using (\ref{eq:AMSBloop}) yields a 2-loop mass for the meson
\begin{equation}\label{eq:sconf_Mmass}
    m_M^2 = \frac{(2\N + 3) \alpha(v)^4 m^2}{(16\pi^2)^2} .
\end{equation}

Along the direction we are considering, this gives a potential
\begin{equation}
    V_{\rm 2-loop} = \frac{(\N+1)(2\N + 3) \alpha(v)^4}{(16\pi^2)^2} m^2 v^2 .
\end{equation}

Notice that at the point (\ref{eq:sconf_chisb}), this is also $\mathcal{O}(m^{2\N/(\N-1)})$. However, since it is 2-loop suppressed, it does not destabilize the chiral symmetry breaking minimum.

We should finally check the effects of higher order terms in the \Kah potential, the leading ones being $(\Tr M^\dagger M)^2$ and $\Tr M^\dagger M M^\dagger M$ with unknown coefficients (including signs). Using (\ref{eq:extendedAMSB}), we find that these give potential terms $\sim m^2 v^4$. At the point (\ref{eq:sconf_chisb}), these are higher order in $m$ and can be neglected.

\subsection{Baryon number breaking direction, $b \neq 0$}

In general one can minimize (\ref{eq:sconf_pot}) with respect to $b$ and $x$, finding
\begin{align}
    b^2 &= \frac{\beta}{\alpha} v^{\N} - 2 x^2 \\
    x &= \frac{(\N - 2) m}{2\alpha} .
\end{align}

Plugging these in we find the runaway potential found in \cite{Luzio:2022ccn}
\begin{equation}\label{eq:sconf_run}
    V|_{b,x} = -\frac{(\N-2)^2 \beta}{2 \alpha} m^2 v^{\N} .
\end{equation}

However, we must account for loop corrections. The bottom $\N$ components of $B$ and $\Bbar$ acquire a mass $\alpha v$, so we integrate them out. This gives, to all but the upper-left component $M_{1 1}$, the 2-loop mass (\ref{eq:sconf_Mmass}). At this point, the remaining superpotential is simply $W = \alpha B_1 M_{1 1} \Bbar_1$. $M_{1 1}$ then obtains a mass at the lower scale $\sqrt{2} \alpha b$. Integrating it out results in 2-loop AMSB masses for $B_1$ and $\Bbar_1$
\begin{equation}
    m_b^2 = \frac{3 \alpha(b)^4 m^2}{(16\pi^2)^2} .
\end{equation}

Adding up these contributions along the direction we are considering, this gives a potential
\begin{equation}\label{eq:sconf_stab}
    V_{\rm 2-loop} = \frac{m^2}{(16\pi^2)^2} [ \N (2\N + 3) \alpha(v)^4 v^2 + 6 \alpha(b)^4 b^2 ] .
\end{equation}
Clearly the first term is dominant. Importantly however, this is the same order in $m$ as the tree level runaway (\ref{eq:sconf_run}) and lower order in $v$ since $\N > 2$. While it is loop (logarithmically) suppressed, this is a smaller effect than the power suppression of (\ref{eq:sconf_run}). Therefore, around the origin where $v \ll 1$, the loop effects stabilize the tree level runaway!

In this case there is also a tri-linear AMSB term coming from (\ref{eq:AMSBloop_trilin}) that goes as $\sim m b^2 x$ with 1-loop suppression. Like the second term in (\ref{eq:sconf_stab}), this is subdominant. Finally, subleading terms in the \Kah lead to power suppressed potential terms that can be neglected.

What we have shown is remarkable: the chiral symmetry breaking point for small $m$ is stable and the AMSB loops effects play a subleading role. However, when we consider a possible runaway direction, the loops come in to save the day. While we cannot be sure of what happens when the fields are $\mathcal{O}(\Lambda)$, there are no runaways from the origin and the chiral symmetry breaking point stands a good chance of being the global minimum.

\subsection{$\N=2$}

In this case tree-level AMSB vanishes because the superpotential is marginal. Due to the positive 2-loop masses, the meson and baryon fields are pushed to the origin of moduli space, where the theory experiences confinement without chiral symmetry breaking. This does not match expectations of non-SUSY QCD and we expect a different global minimum to emerge in the large SUSY breaking limit. A similar phenomemon was seen for a Standard-Model-like chiral $SU(5)$ gauge theory in~\cite{Bai:2021tgl}.

\section{$\N + 1 < \F \leq 3/2 \N$: Free magnetic phase}

For this range of flavors the SUSY theory is in the free magnetic phase and the IR is described by an $SU(\Ntil)$ ($\Ntil = \F - \N$) gauge theory with quarks and anti-quarks in representations $q_i(\bar{\square},\mathbf{1})$ and $\bar{q}_j(\mathbf{1},\square)$ of the $SU(\F)\times SU(\F)$ flavor group, respectively. Additionally, the magnetic theory has a gauge-singlet meson $M_{ij}$ in the $(\square,\bar{\square})$ of the flavor symmetry. The superpotential is given by
\begin{equation}\label{free_mag_W}
    W = \lambda \Tr q_i M_{i j} \bar{q}_j
\end{equation}
where all fields have already been normalized to have canonical K\"ahler potentials. Importantly, only the deep IR behavior of the theory is specified and we do not have control over the relative strengths of the gauge interaction and the Yukawa interaction $\lambda$ in Eq.~(\ref{free_mag_W}).

The case of the free magnetic phase is very subtle, and so far has not been properly analyzed. In fact, this phase is expected to be beset by baryonic runaway directions, so that no useful information can be obtained. We show that for the majority of the free magnetic phase ($N_c+1< N_f \lesssim 1.43 N_c$) the baryonic runaway directions are lifted, and the chiral symmetry breaking minimum is stable and likely the global minimum of the theory. The analysis itself is quite involved, as one has to examine several branches, which we will present below. 

We proceed by first analyzing the baryonic direction, where the entire dual gauge group is Higgsed. As mentioned, the free magnetic phase for $N_f \lesssim 1.43 N_c$ is free of runaways in this direction. We next exhibit the chiral symmetry breaking minimum along the mesonic direction. Finally, we check the mixed directions, where only some meson VEVs are turned on, to ensure that they contain no runaways.

\subsection{RG analysis and baryonic branches}

In a small neighborhood of the origin of moduli space, the theory is allowed to run into the deep IR. As suggested by the name, the theory is IR free, with both the gauge coupling $g$ and Yukawa coupling running to zero. However, their coupled beta functions make them run asymptotically to the IR attractor given by
\begin{equation}\label{eq:match}
    0 = \frac{d}{d \log{\mu}} \frac{g^2}{\lambda^2} .
\end{equation}

This allows $\lambda$ to be written in terms of $g$, and we can use (\ref{eq:AMSBloop}) to find the 2-loop masses of the dual squarks and the mesons
\begin{align}
    m_q^2 &= \frac{(-\widetilde{b}) g^4}{(16\pi^2)^2} \frac{\F^2 - 3\F \Ntil - \Ntil^2 + 1}{2\F + \Ntil} m^2 \label{eq:m_q}\\
    m_M^2 &= \frac{(-\widetilde{b}) \Ntil \lambda^2 g^2}{(16\pi^2)^2} m^2 \label{eq:m_M}
\end{align}
where $\widetilde{b} = 3\Ntil - \F$ is negative. The mesons maintain a positive mass throughout the free magnetic window, as do the dual squarks for most of the window. However, at the upper end $\F \gtrsim 1.43 \N$ (in the large $\N$ limit), the dual squark mass turns negative and we expect a baryonic runaway towards an uncalculable minimum.

Concretely, for $\F \gtrsim 1.43 \N$ we consider giving D-flat VEVs to the dual squark
\begin{equation}
    q = \widetilde{B} \left(\begin{array}{c} 
		1_{\Ntil \times \Ntil} \\
		0_{\Ntil \times \N} \\
		\end{array} \right) .
\end{equation}
The effect of this is to Higgs the dual gauge group at the scale $\widetilde{B}$, and to give masses to the dual anti-quarks and some of the mesons. Substituting their equations of motion eliminates the superpotential. Eq.~(\ref{eq:m_q}) then translates into a tachyonic mass for $\widetilde{B}$, where the gauge coupling is evaluated at the scale $\widetilde{B}$.

The first detailed exploration of baryonic runaways with SUSY breaking applied consistently between the UV and IR was undertaken in~\cite{Cheng:1998xg} (see also the more recent~\cite{Abel:2011wv}). In both of these works, which used different mechanisms to break SUSY, baryonic runaways were present throughout the free-magnetic phase. It is encouraging that AMSB, while not eliminating them, lifts these directions for most of the phase.

\subsection{Mesonic branch}

In this section we give the meson a VEV with full rank, repeating the analysis of \cite{Murayama:2021xfj}. This gives masses to the dual quarks and anti-quarks. Without their effects, the beta function of the gauge theory flips sign, allowing the theory to generate a new IR dynamical scale given by
\begin{equation}
    \Lambda_L^{3\Ntil} = \Lambdatil^{3\Ntil - \F} \det{M} .
\end{equation}
The usual superpotential of pure SYM is generated:
\begin{equation}\label{free_mag_pure}
    W = \Ntil \Lambda_L^3 = \Ntil ( \det M )^{1/\Ntil}
\end{equation}
where as usual we have set $\Lambdatil = 1$. Upon adding tree level AMSB, the minimum can be found along the homogeneous direction $M = v \mathbf{1}$ with the potential
\begin{equation}
    V = \F | v^{\F/\Ntil - 1} |^2 + (\F - 3\Ntil) m v^{\F/\Ntil} + c.c.
\end{equation}
at the point
\begin{equation}
    v = \left( \frac{(3\Ntil-\F)m}{\F-\Ntil} \right)^\frac{\Ntil}{\F-2\Ntil} , V_{\rm min} = -\mathcal{O}\left(m^{2\frac{\F-\Ntil}{\F-2\Ntil}}\right) .
\end{equation}

The 2-loop potential from (\ref{eq:m_M}) contributes at the same order in $m$, however it is loop suppressed. We find that the chiral symmetry breaking minimum is stable.

\subsection{Mixed branches}

Instead of turning on all of the meson VEVs, we can choose to turn on only some of them. These will reveal tree level AMSB contributions within the free magnetic phase with tree level runaways. However, as in the case of s-confinement, the AMSB loop effects will stabilize these directions.

We begin by writing the meson matrix as
\begin{equation}
    M = \left(\begin{array}{cc} 
		\Mtil_{R_f \times R_f} & 0\\
		0 & \widehat{M}_{(\F-R_f) \times (\F-R_f)}\\
		\end{array} \right)
\end{equation}
and without loss of generality we look for minima at diagonal $M$. We then give the lower component $\widehat{M}$ a VEV. This gives masses to $\F-R_f$ flavors of quarks, leaving an $SU(\Ntil)$ gauge theory with $R_f$ massless flavors and a new dynamical scale
\begin{equation}
    \Lambda_L^{3\Ntil-R_f} = \widetilde{\Lambda}^{3\Ntil - \F} \det{\widehat{M}}
\end{equation}
with $\widetilde{\Lambda}$ the Landau pole of the dual theory. In what follows we will set $\widetilde{\Lambda} = 1$. Finally, we assume that $\Mtil$ remains small compared to both $\widehat{M}$ and the generated scale $\Lambda_L$.

For $1 \leq R_f < \Ntil$, the remaining theory is of ADS-type and has the superpotential
\begin{equation}
    W = (\Ntil - R_f) \left( \frac{\Lambda_L^{3\Ntil - R_f}}{\det{N}} \right)^\frac{1}{\Ntil-R_f} + \Tr{\Mtil N}
\end{equation}
where $N$ is the meson formed by the remaining massless dual-quarks. We have ignored $\lambda$ as it will be irrelevant for this discussion. The second term comes from the Yukawa of the dual theory.

The SUSY equation of motion (EOM) for $\Mtil$ sets $N = 0$. Evidently, the EOM for $N$ is singular at this point and to compensate we must have $\Mtil \rightarrow \infty$. However, this violates the assumption of small $\Mtil$. Therefore, even before a small AMSB deformation can be applied, this branch collapses back to the mesonic branch already considered.

Next consider the case of $R_f = \Ntil$, which will have emergent meson and baryon degrees of freedom with a quantum modified constraint. Furthermore, the superpotential $W = \Tr{\Mtil N}$ fixes $\Mtil = N = 0$. We thus find ourselves at the baryon point where as before the baryons are stable, but this time with the emergent meson directions stabilized by a superpotential. The only question that remains is the $\widehat{M}$ dependence. For simplicity consider $\widehat{M} = v \mathbf{1}$. The new dynamics will generate at leading order the \Kah potential term
\begin{equation}
    K \supset a \Lambda_L^2 = a v^{2C}
\end{equation}
where $a$ is an $\mathcal{O}(1)$ number of unknown sign and $C = (\F - R_f)/(3\Ntil - R_f) > 1$. This will give rise to a tree level AMSB potential of $\mathcal{O}(m^2 v^{2C})$. However, as before the 2-loop AMSB mass for the meson will give a positive contribution at $\mathcal{O}(m^2 v^2)$, stabilizing this direction.

For $\Ntil + 1 \leq R_f < 3\Ntil$, the IR dynamics of the remaining theory are described by a magnetic dual with gauge group $SU(R_f-\Ntil)$ (except for $R_f=\Ntil+1$ where the theory is s-confining). The superpotential is
\begin{equation}
    W_L = \Tr b_i N_{i j} \bar{b}_j + \Tr{\Mtil N} .
\end{equation}
The $N$, $b$, and $\bbar$ are dual mesons, quarks (baryons), and anti-quarks (anti-baryons) formed by the massless dual quarks. Again the superpotential term (\ref{free_mag_W}) has transformed to enforce $N=0$ in the supersymmetric limit. This means when we introduce tree-level AMSB, $N=\mathcal{O}(m)$, and we were justified in ignoring the s-confining $\det{N}$ term as a high power of $m$ (assuming $N$ is even full rank). We rescale the fields by appropriate factors of $\Lambda_L$ to make them canonical. Ignoring order one factors we have
\begin{equation}
    W_L = \Tr b_i N_{i j} \bar{b}_j + \Lambda_L \Tr{\Mtil N} .
\end{equation}
Finally we substitute the value of $\Lambda_L$ (and set $\widetilde{\Lambda} = 1$) to arrive at
\begin{equation}
    W_L = \Tr b_i N_{i j} \bar{b}_j + (\det{\widehat{M}})^{1/(3\Ntil-R_f)} \Tr{\Mtil N} .
\end{equation}

Let all fields be real and consider the direction given by $N_{ii} = n_i$, $\Mtil_{ii} = x_i$, $b_{ii} = -\bbar_{ii} = y_i$, for $i = 1,\dots,(R_f-\Ntil)$ and with all other entries $0$. Finally let $\widehat{M} = v \mathbf{1}$.

The potential is
\begin{align}
    V &= \sum_i \left( 2y_i^2 n_i^2 + (v^C x_i - y_i^2)^2 + v^{2C} n_i^2 \right. \nonumber\\
    &\quad \left. + 2(C-1)m v^C n_i x_i \right) + \frac{C}{3\Ntil-R_f} v^{2C-2} \left( \sum_i n_i x_i \right)^2
\end{align}
where $C$ is defined as before and remains greater than $1$. Notice that the final term is smaller than the third term in the first sum by a factor of $x^2/v^2 \ll 1$. Therefore, we can neglect this term and the potential splits into $R_f-\Ntil$ identical parts. In what follows, we suppress the index $i$. Substituting the $y$ and $n$ equations of motion, and using $n, x \ll \Lambda_L = v^C$ along the way, we get
\begin{equation}
    V|_{y,n} = -(C-1)^2 m^2 x^2 .
\end{equation}

As long  we keep $x \ll v^C$, we can let $x, v \rightarrow 1$, signalling a tree level minimum of $-\mathcal{O}(m^2)$ in the incalculable region where field VEVs are $\mathcal{O}(\Lambda)$. Note that in this direction all fields, baryonic and mesonic, are turned on.

However, as we saw for the s-confining runaway, the loop effects must be considered. While this tree-level runaway is power suppressed as $\mathcal{O}(x^2) \ll \mathcal{O}(v^{2C})$, the 2-loop potential gives a positive contribution with $\mathcal{O}(v^2)$. Therefore, there is again no runaway.

When $R_f \geq 3\Ntil$, the theory remains IR free and there are no tree level runaways. As long as $\F \lesssim 1.43 \N$ the dual quarks will have positive 2-loop AMSB mass.

In summary, we have demonstrated that there is a stable chiral symmetry breaking minimum and that for $\F \lesssim 1.43 \N$ there are no runaways.

\section{$3/2 \N < \F < 3 \N$: Conformal window} 

In the conformal window, the magnetic description is no longer IR free. Rather, it has a non-trivial fixed point, which is weakly coupled at the lower end of the window. We will first analyze the behavior of AMSB in this region and find baryonic runaways to incalculable minima. Then, we will turn to the upper end of the window where the electric theory has a weakly coupled fixed point. As demonstrated in \cite{Murayama:2021bndrv}, AMSB makes a relevant deformation and destroys the superconformal phase. We can only conjecture about the intermediate region where both descriptions are strongly coupled. Finally, we demonstrate local chiral symmetry breaking minima throughout the window.

\subsection{Lower conformal window}

We begin by considering $\F = 3\Ntil/(1+\epsilon)$ where $\epsilon \ll 1$, and will work in the large $\Ntil$ limit and leading non-trivial order in $\epsilon$ for simplicity. For notational convenience, we define
\begin{align}
	x &\equiv \frac{\tilde{N}_c}{8\pi^2} \lambda^2, \qquad
	y \equiv \frac{\tilde{N}_c}{8\pi^2} g^2.
\end{align}

The beta functions of the magnetic theory, including the 2-loop contribution for $y$, are
\begin{align}
	\beta(x) &= x ( -2y + 7x ), \label{eq:RGEg} \\
	\beta(y) &= -3y^2 ( \epsilon - y + 3x) . \label{eq:RGElambda}
\end{align}
They admit a BZ fixed point at $(x_0,y_0) = (2\epsilon,7\epsilon)$. As the theory flows to the IR, $x$ and $y$ will approach this point from above, along the trajectory specified by (\ref{eq:match}). Define $\delta x = x - x_0$ and $\delta y = y - y_0$. Close to the fixed point this yields
\begin{equation}
    \delta x = \frac{2}{7}\left(1 + \frac{3}{2} \epsilon \right) \delta y .
\end{equation}
The RG flow is
\begin{equation}
    \beta(y) = 21 \epsilon^2 \delta y
\end{equation}
yielding
\begin{equation}
    \delta y \sim \mu^{21\epsilon^2} .
\end{equation}

Using (\ref{eq:AMSBloop}), the meson and dual squark masses are
\begin{align}
    m_M^2 &= \frac{3}{2}\epsilon^2 \delta y \, m^2 \\
    m_q^2 &= -\frac{3}{4}\epsilon^2 \delta y \, m^2 .
\end{align}
Thus in the lower conformal window the dual squarks are tachyonic and there is a runaway to an incalculable minimum.

\subsection{Upper conformal window}

We now examine the upper conformal window via the electric description, reviewing the results of \cite{Murayama:2021bndrv}. Now $\F = 3\N/(1+\epsilon)$, and we use all conventions of the previous section. The beta function at 2-loop is
\begin{equation}
	\beta(y) = -3y^2 ( \epsilon - y)
\end{equation}
where the BZ fixed point $y_0 = \epsilon$ is now approached from below as
\begin{equation}
    (-\delta y) \sim \mu^{3\epsilon^2} .
\end{equation}

From (\ref{eq:AMSBloop}) and (\ref{eq:AMSBloop_glu}) we obtain the squark and gluino masses
\begin{align}
    m_Q^2 &= \frac{3}{4} \epsilon^2 (-\delta y) m^2 \\
    m_\lambda &= \frac{3}{2} (-\delta y) m .
\end{align}

As expected the squark mass is positive. As long as $3\epsilon^2 < 1$ (this bound is outside of our small $\epsilon$ limit and should be taken with a grain of salt), at some point in the RG flow the squark and gluino masses will exceed the renormalization scale. At this point the superpartners can be integrated out and the superconformal phase is destroyed. What remains is non-SUSY QCD and must be analyzed from the (albeit strongly-coupled) magnetic description.

\subsection{Chiral symmetry breaking minimum}

We have shown that AMSB, at both the top and bottom of the conformal window, destroys the superconformal phase. It is reasonable to assume this is the case throughout the window. Furthermore, we demonstrated that at the bottom of the window the theory has a runaway to an incalculable minimum.

Looking instead for local minima, we examine the mesonic branch. Just as in the free magnetic phase, this gives masses to the dual quarks and generates a new dynamical scale. The superpotential is given by (\ref{free_mag_pure}). However, unlike the free magnetic phase where the \Kah receives logarithmic wave-function renormalization (which we ignored), in the conformal window we have
\begin{align}
	Z_M (\mu) \sim \mu^{1 - 3\Ntil/\F}
\end{align}
which is evaluated at $\mu = v$, where $M = v \mathbf{1}$. The result is that the scaling of the local chiral symmetry minimum is modified to \cite{Murayama:2021bndrv}
\begin{align}
	V = -\mathcal{O}( m^\sigma), \quad
	\sigma = 1 + \frac{\F^2}{\F^2 - 3\F \Ntil + 3 \Ntil^2} .
\end{align}
Note that $\sigma$ goes from $4$ ($\F=\frac{3}{2}N_c$) to $5$ ($\F =2 N_c$) back to $4$ ($\F = 3 N_c$).

\section{$\F \geq 3\N$: Free electric phase}

For large number of flavors, the 2-loop squark mass from AMSB is negative, leading to true runaway behavior. AMSB cannot be used to understand the non-SUSY theory in this case.

\section{Conclusions}

We carefully analyzed the behavior of $SU(\N)$ gauge theories with $\F$ flavors upon the application of AMSB, focusing on the chiral symmetry breaking minima and potential baryonic runaway directions. For $\N+1 \leq \F \leq 3/2 \N$ we found that naive tree level runaways are power suppressed in comparison to loop effects, which stabilize these directions. However, a true loop level runaway was found for the upper end of the free magnetic phase, $\F \gtrsim 1.43 \N$. This baryonic runaway continued into the lower end of the conformal window, and we cannot discount such runaways throughout the window. Such runaways point to the existence of some non-calculable minimum at large field values of ${\cal O} (\Lambda )$, which may or may not correspond to the global minimum of the theory. 

The case of $\F = \N$ required particular care due to the inherently strongly coupled nature of the quantum modified moduli space. We found that the theory is best analysed after implementing the quantum constraint. Upon application of AMSB the stability of the chiral symmetry breaking point cannot be determined. This is not due to a problem with the AMSB method, but rather because the \Kah potential terms that are critical to this determination are incalculable. 

In summary we found (with the exception of the cases $\F = \N$ for $\N > 2$ and $\F = \N + 1$ for $\N=2$) that stable chiral symmetry breaking minima are present for $\F < 3 N_c$ upon application of AMSB in the small SUSY-breaking limit. Furthermore, the theories with $\F \lesssim 1.43 \N$ are protected from runaways to incalculable minima. This does not prove that there are no deeper minima with fields of $\mathcal{O}(\Lambda)$, however we take it to be strong evidence for the conjecture that in these cases the chiral symmetry breaking minima are in fact global.

Our analysis was performed in the $m \ll \Lambda$ limit, and the question remains about the behavior in the non-supersymmetric limit of $m \gg \Lambda$. The existence of the chiral symmetry breaking minima for all flavors is indicative that these are continuously connected to the true vacua of non-SUSY QCD. Irrespective of the potential appearance of a phase transition between these two limits (see arguments based on holomorphy in~\cite{Csaki:2021xhi,Csaki:2021aqv}, and also see~\cite{Dine:2022req,Dine:2022nmt}), these are the vacua that are of phenomenological interest for the study of real-world QCD.


\begin{acknowledgments}

We are grateful to Nathaniel Craig for emphasizing the role of potential baryonic runaway directions and for numerous discussions about them, and about the AMSB method in general. CC is supported in part by the NSF grant PHY-2014071 as well as the US-Israeli BSF grant 2016153. AG is supported in part by the NSERC PGS-D fellowship, and in part by the NSF grant PHY-2014071. OT and HM were supported in part by the DOE under grant DE-AC02-05CH11231. HM is the Hamamatsu Professor at the Kavli IPMU in Tokyo. 
HM was also supported in part by the NSF grant
PHY-1915314, by the US-Israeli BSF grant 2018140, by the JSPS Grant-in-Aid for
Scientific Research JP20K03942, MEXT Grant-in-Aid for Transformative Research Areas (A)
JP20H05850, JP20A203, by WPI, MEXT, Japan, and Hamamatsu Photonics, K.K.

\end{acknowledgments}

\bibliographystyle{utcaps_mod}
\bibliography{AMSBQCD}

\end{document}